\documentclass[twocolumn,showpacs,prl]{revtex4}
\usepackage{graphicx}
\usepackage{dcolumn}
\usepackage{bm}

\begin{document}

\title{Trapped ion quantum computation with transverse phonon modes}
\author{Shi-Liang Zhu, C. Monroe, and L.-M. Duan}

\address{FOCUS Center and MCTP,
Department of Physics, University of Michigan, Ann Arbor, MI 48109}

\begin{abstract}
We propose a scheme to implement quantum gates on any pair of trapped ions
immersed in a large linear crystal, using interaction mediated by the
transverse phonon modes. Compared with the conventional approaches based on
the longitudinal phonon modes, this scheme is much less sensitive to ion heating and thermal motion outside of the Lamb-Dicke limit thanks to the
stronger confinement in the transverse direction. The cost for such a gain
is only a moderate increase of the laser power to achieve the same gate
speed. We also show how to realize arbitrary-speed quantum gates with
transverse phonon modes based on simple shaping of the laser pulses.
\end{abstract}


\maketitle

Trapped ions have been demonstrated as one of the most promising
systems for implementation of quantum computation. Different
theoretical schemes have been proposed for quantum gate operations
\cite{Monroe2002,Cirac1995,Soresen1999,Milburn,f1,f2,Zhu2005}, and
many building blocks of quantum computing have been experimentally
demonstrated
\cite{Wineland,Monroe95,Sackett,Liebfried,Gulde,Blinov,Barrett,Haljan}.
In an ion trap quantum computer, entangling gates between different ions are mediated through phonons in the collective ion motion. In all previous gate schemes
\cite{Cirac1995,Soresen1999,Milburn,f1,f2,Zhu2005}, the longitudinal phonon (LP) modes are exploited by kicking the ions along the axial direction of a linear trap.

In this work, we propose to use the transverse phonon (TP) modes for
gate operations. Compared with the conventional schemes (hereafter referred to as LP
gates), gates involving TP modes (TP\ gates) have
the following distinctive features: First, due to the strong confinement
in the transverse direction, the TP gate is much less sensitive to ion
heating and thermal motion. Even if the axial ion oscillation amplitude
is significantly greater than the optical wavelength
(outside of the Lamb-Dicke regime), high-fidelity gates through the TP modes are still possible.
If $\beta$ denotes the ratio
of the center-of-mass (CM) trap frequencies for the transverse and the longitudinal directions ($%
\beta \gg 1$ in typical experiments), we show that gate infidelity
due to thermal ion motion is reduced by a factor ranging from $\beta ^{4}$ to $%
\beta ^{6}$, depending on details of the initial ion temperature and
the heating mechanism. This improvement may be particularly
significant for a system of many ions or for ions confined in a
microtrap \cite{Stick}, where ion heating and thermal motion may
dominate gate erors. Second, the cost of using the TP modes is
moderate, even though it is more difficult to excite the TP modes
due to their strong confinement. For TP gates to have the same speed
as LP gates, the intensity of the driving laser needs only to be
increased by a factor of $\sqrt{\beta/2 }$, a small factor when
compared with the improvement in gate fidelity. Finally, we show
that TP quantum gates can be operated with arbitrary speeds.
Although the frequency splitting of the TP modes is significantly
smaller than that of the LP modes, this does not impose any limit to
the gate speed. High-fidelity  fast TP gates are still possible
through control of a simple sequence of laser pulses, which
typically involves excitation of many TP modes.

\textbf{Transverse phonon (TP) modes.} To design TP quantum gates,
we first describe the structure of the TP modes. We consider a
system of $N$ ions confined in a linear trap. The phonon modes are
obtained through diagonalization of the Hamiltonian for the ion
external motion \cite{James}. The ion motional Hamiltonian has the
form $H_{0}=\sum_{\xi }\sum_{j=1}^{N}\frac{p_{\xi ,j}^{2}}{2M}+V,$
where $M$ is the ion mass, and $p_{\xi ,j}$ is the momentum operator
of $j$th ion along the
direction $\xi $ ($\xi =x,y,z$). The potential $V=\frac{M}{2}%
\sum_{j=1}^{N}\left( \omega _{x}^{2}x_{j}^{2}+\omega
_{y}^{2}y_{j}^{2}+\omega _{z}^{2}z_{j}^{2}\right) +\sum_{n,j}\frac{e^{2}}{%
4\pi \epsilon _{0}r_{nj}}$ accounts for the Coulomb interaction
between the ions as well as the external trapping potential, where
$r_{nj}$ denotes the distance between ions $n$ and $j$, and $\omega
_{\xi }$ is the CM trap frequency along the direction $\xi $.
Typically, $\omega _{x}\sim \omega _{y}\gg \omega _{z}$, and one has
a linear geometry with an ion chain along the $\mathbf{z}$ axis when
$\omega _{x,y}/\omega _{z}$ is larger than the
critical ratio about $0.73N^{0.86}$\cite{Steane}. The equilibrium positions $%
z_{n}^{0}$ ($x_{n}^{0}=y_{n}^{0}=0$) of the ions are determined by
the condition $[\partial V/\partial z_{n}]_{z_{n}=z_{n}^{0}}=0,$
$\left( 1\leq n\leq N\right) $. With the dimensionless parameters
$u_{n}\equiv z_{n}^{0}/\ell $ $(n=1,2,\dots ,N),$ where
$\ell \equiv \sqrt[3]{e^{2}/4\pi \epsilon _{0}M\omega _{z}^{2}}$ sets the scale of ion spacings in the linear crystal, this condition yields a set of equations $%
u_{j}-\sum_{n=1}^{j-1}1/(u_{j}-u_{n})^{2}+%
\sum_{n=j+1}^{N}1/(u_{j}-u_{n})^{2}=0,$ which can be solved numerically to
determine $z_{n}^{0}$ for any larger ion array \cite{James}. To find all
the phonon modes, one then just needs to expand the potential $V$ around the
ions' equilibrium positions $\xi _{n}^{(0)}\equiv \left( x_{n}^{0},\text{ }%
y_{n}^{0},\text{ }z_{n}^{0}\right) $ with $\xi _{n}=\xi
_{n}^{(0)}+q_{n}^{\xi }$. Under the harmonic approximation, we have $%
V=\left( 1/2\right) \sum_{\xi ,n,j}q_{n}^{\xi }q_{j}^{\xi }[\partial
^{2}V/\partial \xi _{n}\partial \xi _{j}]_{\xi _{n}=\xi _{n}^{0}}=\left(
M\omega _{z}^{2}/2\right) \sum_{\xi ,n,j}A_{nj}^{\xi }q_{i}^{\xi }q_{j}^{\xi
}$, where the matrix elements
\begin{equation}
A_{nj}^{\xi }=\left\{
\begin{array}{ll}
\beta _{\xi }^{2}+\sum\limits_{p=1,p\not=j}^{N}a_{\xi }/|u_{j}-u_{p}|^{3} &
(n=j) \\
-a_{\xi }/|u_{j}-u_{n}|^{3} & (n\not=j),
\end{array}
\right.
\end{equation}
with $\beta _{\xi }=\omega _{\xi }/\omega _{z}$, $a_{x}=a_{y}=-1$, and $%
a_{z}=2$ \cite{Deng}. The eigenfrequencies $\omega _{\xi ,k}\equiv \sqrt{\lambda _{\xi ,k}}%
\omega _{\xi }$ and eigenvectors $\mathbf{b}_{j}^{\xi ,k}$ of the
normal phonon modes are obtained from diagonalization of the matrix
$A^{\xi }=[A_{nj}^{\xi }]$ with $\sum_{n}A_{nj}^{\xi
}\mathbf{b}_{n}^{\xi ,k}=\lambda _{\xi ,k}\mathbf{b}_{j}^{\xi ,k}$.
Then, with the substitution $q_{j}^{\xi
}(t)=\sum_{k}\mathbf{b}_{j}^{\xi ,k}\sqrt{\hbar /2M\omega _{\xi
,k}}(a_{\xi ,k}+a_{\xi ,k}^{\dagger })$, the motional Hamiltonian
$H_{0}$ reduces to the standard form $H_{0}=\sum_{\xi
}\sum_{k=1}^{N}\hbar \omega _{\xi ,k}(a_{\xi ,k}^{\dagger }a_{\xi
,k}+1/2)$, expressed by the annihilation and creation operators
$a_{\xi ,k},a_{\xi ,k}^{\dagger }$ of the $k$th normal mode in the
$\xi $ direction.

In order to visualize the TP and LP modes, we plot the complete mode
spectrum for a $10$-ion array in Fig. 1 (the modes along $x$ and $y$
directions are degenerate, so it is enough to show only the
$x$-modes). We choose the trap frequency ratio $\beta _{x}=10$,
which is typical for experiments and larger than the critical value
of $5.3$ to stabilize a linear configuration for $N=10$ ions. As
opposed to the LP modes, the highest frequency TP mode is the
center-of-mass mode at $\omega _{x}$. The frequency splitting
between the CM\ mode and the second-to-highest mode (the bending
mode) is about $0.05\omega _{z}$ for $N=10$, which is significantly
smaller than the splitting $(\sqrt{3}-1)\omega _{z}$ of the
corresponding LP modes (the spectral structure of the TP modes is
inverted compared to the LP modes, as seen in Fig. 1).
For entangling local ions (such as neighboring ions), it is best to use the low-frequency TP ``zigzag" mode \cite{Schiffer} as it is more resolved from the other TP modes and most insensitive to the ion
heating. But the CM\ mode has the advantage that it is equally coupled to
all the ions, and thus more appropriate for gates between nonlocal ions (such as ions at different edges of the chain).
 When comparing features of gates using TP or LP modes, we parameterize the comparision of  the CM\ modes for both cases.

\begin{figure}[tbph]
\label{Fig1} \includegraphics[height=3cm,width=8cm]{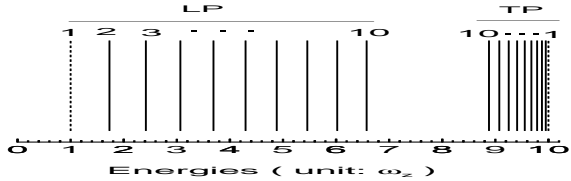}
\caption{The spectrum of the longitudinal (LP) and transverse phonon
(TP) modes for a 10-ion array. The dashed lines (No. 1) denote the
center of mass (CM) mode.}
\end{figure}

\textbf{General formalism of trapped ion quantum gates.} First, we give a general
formalism for multi-ion entangling quantum gates, which is valid with either TP\ or
LP\ modes. The qubit for each ion is represented by two hyperfine
states, denoted as $|0\rangle $ and $|1\rangle $ in general. The gate is
achieved by applying a state-dependent ac-Stark shift on the ions \cite
{Liebfried,Haljan}, using two laser beams of equal intensity, wavevector difference $\Delta \mathbf{k}$ and frequency difference $\mu $.
As it is common in experiments, we assume that the average
ac-Stark shifts are the same for the $|0\rangle $ and $|1\rangle $ states
for the ions in their equilibrium positions. In this case, the Hamiltonian
for the laser-ion interaction has the form
\begin{equation}
H=\sum_{j=1}^{N}\hbar \Omega _{j}\cos (\Delta \mathbf{k}\cdot \mathbf{q}%
_{j}+\mu t)\sigma _{j}^{z},  \label{H}
\end{equation}
where $\Delta \mathbf{k\cdot q}_{j}=\sum_{\xi }\Delta \mathbf{k}_{\xi
}q_{j}^{\xi }(t)$, and $\Omega _{j}$ denotes the two-photon Rabi frequency
of the $j$th ion, which is proportional to the intensity of the driving laser.
For convenience, $\Omega _{j}$ is assumed to be real, but it can be
time-dependent.

Now we assume that the relative wave vector $\Delta \mathbf{k}$ is
chosen along a certain direction $\xi $ ($\xi =x$ or $z$), and
motion in all modes in this direction is in the Lamb-Dicke regime
$\eta _{\xi ,k}\sqrt{\bar{n}_{\xi ,k}+1}\ll 1$ for all $k$, where
$\eta _{\xi ,k}=\left| \Delta \mathbf{k}\right| \sqrt{\hbar/2M\omega
_{\xi ,k}}$ is the Lamb-Dicke parameter and $\bar{n}_{\xi ,k}$ the
mean phonon occupation number of mode $(\xi ,k)$. Note that for TP
quantum gates ($\xi =x$), the lower frequency LP modes $\xi =z$ (as
well as the other transverse mode $\xi =y$) are decoupled and hence
need not be confined within the Lamb-Dicke regime.  To lowest order
in $\eta _{\xi ,k}$ and under the rotation-wave approximation, the
interaction-picture Hamiltonian of the system is
\begin{equation}
H_{I}=-\sum_{j,k=1}^{N}\hbar \chi_{j}^{\xi }(t)g_{\xi ,j}^{k}(a_{\xi
,k}^{\dagger }e^{i\omega _{\xi ,k}t}+a_{\xi ,k}e^{-i\omega _{\xi
,k}t})\sigma _{j}^{z},  \label{H_int}
\end{equation}
where the coupling constant $g_{\xi ,j}^{k}=\eta _{\xi ,k}\mathbf{b}%
_{j}^{\xi ,k}$, and $\chi_{j}(t)=\Omega _{j}\sin (\mu t)$ is proportional to the
state-dependent force on the $j$-th ion.

The evolution operator corresponding to the Hamiltonian $H_{I}$ is given by
\cite{Zhu2003, Zhu2005}
\begin{equation}
U(\tau )=\exp [i\sum_{j}\phi _{j}^{\xi }(\tau )\sigma
_{j}^{z}+i\sum_{j<n}\phi _{jn}^{\xi }(\tau )\sigma _{j}^{z}\sigma _{n}^{z}],
\label{U2}
\end{equation}
where the displacement operator $\phi _{j}^{\xi }(\tau )=\sum_{k}[\alpha
_{\xi ,j}^{k}(\tau )a_{\xi ,k}^{\dagger }+\alpha _{\xi ,j}^{k\ast }(\tau
)a_{\xi ,k}]$ with $\alpha _{\xi ,j}^{k}(\tau )=\int_{0}^{\tau
}\chi_{j}(t)g_{\xi ,j}^{k}e^{i\omega _{\xi ,k}t}dt$, and the conditional phase $%
\phi _{jn}^{\xi }(\tau )=2\int_{0}^{\tau
}dt_{2}\int_{0}^{t_{2}}dt_{1}\sum_{k}\chi_{j}(t_{2})g_{\xi ,j}^{k}g_{\xi
,n}^{k}\chi_{n}(t_{1})\sin \omega _{\xi ,k}(t_{2}-t_{1}).$ In the case of a
time-independent $\Omega _{j}$, $\alpha _{\xi ,j}^{k}(\tau )$ and $\phi
_{jn}^{\xi }(\tau )$ have the following explicit expressions:

\begin{equation}
\alpha _{\xi ,j}^{k}=\frac{\Omega _{j}g_{\xi ,j}^{k}\left\{ \mu +e^{i\omega
_{\xi ,k}\tau }[-\mu \cos (\mu \tau )+i\omega _{\xi ,k}\sin (\mu \tau
)]\right\} }{(\mu ^{2}-\omega _{k}^{2})},  \label{Alpha}
\end{equation}

\[
\phi _{jn}^{\xi }=2\Omega _{j}\Omega _{n}\sum_{k}\frac{g_{\xi ,j}^{k}g_{\xi
,n}^{k}}{\mu ^{2}-\omega _{\xi ,k}^{2}}\left\{ \frac{\omega _{\xi ,k}[-2\mu
\tau +\sin (2\mu \tau )]}{4\mu }+\right.
\]
\begin{equation}
\left. \frac{\mu \lbrack \omega _{\xi ,k}\cos (\omega _{\xi ,k}\tau )\sin
(\mu \tau )-\mu \cos (\mu \tau )\sin (\omega _{\xi ,k}\tau )]}{\mu
^{2}-\omega _{\xi ,k}^{2}}\right\} .  \label{Phase}
\end{equation}

To drive a conditional phase flip (CPF) gate between arbitrary ions
$j$ and $n$, we take $\Omega _{j}$ to be nonzero only for these two
ions, and set the laser detuning $\mu $ and the gate time $\tau $ so
that $\phi _{j}^{\xi }(\tau )=\phi _{n}^{\xi }(\tau )=0$ and $\phi
_{jn}^{\xi }(\tau )=\pi /4$. In this case, the evolution operator
$U(\tau )$ reduces to the CPF operator $U_{jn}=\exp (i\pi \sigma
_{j}^{z}\sigma _{n}^{z}/4)$.

\textbf{TP versus LP quantum gates.} To compare quantum
gates based on TP vs. LP modes, let us start with the
assumption that the driving laser can address individual phonon modes
through frequency selection (resolved-sideband addressing). This requires the
gate time $\tau $ to be much larger than $\tau _{\Delta }\equiv 2\pi /\Delta
$, where $\Delta $ is the characteristic frequency splitting of the phonon
modes (we will see later that only one phonon mode dominates when
$\tau \geq 2\tau _{\Delta }$). The sideband addressing assumption, although
not essential, allows us to derive simple
analytic relations that permit a direct comparison between TP and LP gates.

With sideband addressing, we dominantly excite a particular phonon mode $%
\left( \xi ,p\right) $ with frequency $\omega _{\xi ,p}$ by adjusting the
laser detuning $\mu $ close to $\omega _{\xi ,p}$ with $\mu =\omega _{\xi
,p}+2\pi l_{\xi }/\tau $, where $l_{\xi }$ is an integer, typically chosen
as $1$ or $-1$ \cite{Soresen1999,Zhu2005}. The qubit state and the motion
state should be disentangled at the end of the gate, which requires $\phi
_{j}^{\xi }(\tau )=\phi _{n}^{\xi }(\tau )=0$ and thus $\tau =l_{\xi
}^{\prime }\pi /\omega _{\xi ,p}$ (see Eq. (4)), where $l_{\xi }^{\prime }$
is another integer (typically, $l_{\xi }^{\prime }\gg l_{\xi }$). From Eq.(%
\ref{Phase}), the conditional phase shift is found to be
\begin{equation}
\phi _{jn}^{\xi }(\tau )=-\frac{\mathbf{b}_{\xi ,j}^{p}\mathbf{b}_{\xi
,n}^{p}}{4\pi }\frac{\eta _{\xi ,p}^{2}\Omega ^{2}\tau ^{2}}{1+l_{\xi
}/l_{\xi }^{\prime }}.  \label{Phi_ij}
\end{equation}
The condition $\phi _{jn}^{\xi }(\tau )=\pi /4$ can be satisfied with an
appropriate choice of the Rabi frequency $\Omega $.

To consider inherent infidelity of the gate operation, we note that
all the above results are derived based on the Lamb-Dicke condition.
In practice, the Lamb-Dicke parameter is finite, and the thermal
motion of the ions will induce some fluctuation of the Rabi
frequency $\Omega^{2}$ in Eq. (6) and lead to gate errors. To
estimate this noise, we need to expand the Hamiltonian of Eq.
(\ref{H}) to higher-orders in the Lamb-Dicke parameters. The effect
of these higher-order terms is to replace $\Omega$ in Eq. (6) with
an effective Rabi frequency $\Omega _{n}^{\xi}$ that depends on the
phonon number $n_{\xi }$ of the ion vibrational mode $\left( \xi
,p\right) $. To the next order of the Lamb-Dicke parameter $\eta
_{\xi ,p}$, we find $(\Omega _{n}^{\xi })^{2}\approx \Omega _{\xi
}^{2}[1-\eta _{\xi ,p}^{2}(2n_{\xi }+1)]$ \cite{Soresen1999}. The
gate fidelity $F_{g}^{\xi }$ \cite{Fidelity} is then found to be
$F_{g}^{\xi }=\frac{1}{2}+\frac{1}{2}\sum_{n=0}^{\infty }P_{n}^{\xi
}\cos \left[\frac{\pi}{2} \eta _{\xi ,p}^{2}(2n+1)\right]$
\cite{note}.  When the initial phonon number distribution
$P_{n}^{\xi }$ takes the form of a thermal state, $P_{n}^{\xi
}=\bar{n}_{\xi }^{n}/(1+\bar{n}_{\xi })^{n+1}$, we find to lowest
order in $\eta _{\xi ,p}$ a gate infidelity $F_{in}^{\xi}\equiv
1-F_{g}^{\xi }$ of
\begin{equation}
F_{in}^{\xi }\approx \pi ^{2}\eta _{\xi ,p}^{4}(\bar{n}_{\xi }^{2}+\bar{n}%
_{\xi }+1/8).  \label{Fin}
\end{equation}

As the TP\ mode has a larger vibrational frequency, the TP\ quantum gates
have a significantly smaller gate infidelity from thermal ion
motion. Even if we assume the TP and the LP modes have the same mean phonon
number $\bar{n}_{x}=\bar{n}_{z}$, the infidelity for the TP\ gate is
smaller by a factor of $F_{in}^{x}/F_{in}^{z}=\eta _{x,p}^{4}/\eta
_{z,p}^{4}=\left( \omega _{z,p}/\omega _{x,p}\right) ^{2}$. In
practice, if the TP and LP modes are subject to the same heating mechanism, and
initially prepared with the same laser cooling technique, we further expect $\bar{n}%
_{x}\ll \bar{n}_{z}$ due to differences in the initial temperature and the heating rate of TP vs. LP modes. The temperature limit $T_{L}$ can be considered as independent of the
phonon frequency $\omega _{\xi ,p}$ for Doppler cooling, and is roughly
proportional to $1/\omega _{\xi ,p}$ for the Raman sideband cooling. So the
contribution of $T_{L}$ to the mean phonon number $\bar{n}_{\xi }$,
estimated as $k_{B}T_{L}/\hbar \omega _{\xi ,p}$, is taken to be $(1/\omega _{\xi ,p})^{\gamma }$, where $\gamma $ is between $1$ and $2$. The
ion heating rate $\dot{\bar{n}}_{\xi}$ for
the phonon mode $\left( \xi ,p\right) $ is proportional to the noise power
spectrum $S(\omega _{\xi ,p})$ at the frequency $\omega _{\xi ,p}$ \cite{HeatingData}, taken to be independent of frequency (white noise) or proportional to $1/\omega _{\xi ,p}$ ($1/f$ noise).
For these practical noise sources \cite{HeatingData}, the average phonon occupation number therefore scales as $(1/\omega _{\xi ,p})^{\gamma'}$, with $\gamma'$ again between $1$ and $2$. If we assume the term $\bar{n}_{\xi }^{2}$
dominates in the infidelity expression (8), which is likely for
many ions in a crystal, the infidelity ratio of TP vs. LP gates is
then $F_{in}^{x}/F_{in}^{z}\sim \bar{n}_{x}^{2}\eta _{x,p}^{4}/\left( \bar{n}%
_{z}^{2}\eta _{z,p}^{4}\right) \sim \left( \omega _{z,p}/\omega
_{x,p}\right) ^{2+\gamma+\gamma'}$, where we have assumed that the gate time $\tau$ is the same for both cases. For the CM\ modes, $\omega _{z,p}/\omega _{x,p}$ is given by the trap frequency ratio $\beta _{x}=\omega _{z}/\omega _{x}$.
So, compared with LP gate, the inherent infidelity of the TP\ gate could be
reduced by a factor of $\beta _{x}^{4}$ to $\beta _{x}^{6}$, which is
significant even for a moderate trap frequency ratio of $\beta _{x}\sim 5$.

Now we look at the cost of the TP quantum gate. As the TP modes have a higher
vibrational frequency, it is harder to excite them, and we need more laser
intensity for the same gate speed. From Eq. (7), to have the same gate time $%
\tau $, the ratio of the required laser intensity $I_{x}/I_{z}$ (note that $%
I_{\xi }$ is proportional to the two-photon Rabi frequency $\Omega _{\xi }$)
is given by $I_{x}/I_{z}\sim \eta _{z,p}/\eta _{x,p}=\sqrt{\omega
_{x,p}/\omega _{z,p}}=\sqrt{\beta _{x}}$ (we neglect $l_{\xi }/l_{\xi
}^{\prime }$ in Eq. (7) as it is typically much less than $1$)$.$ So,
although we need additional laser power for the TP quantum gate, this cost
is moderate compared with the improvement we achieve in the gate
fidelity. If we take into account of different laser excitation
configurations for the TP and the LP gates, this cost is even less. For the
LP gate, the relative wave vector $\Delta \mathbf{k}$ of the Raman laser
beams needs to be along the trap axis, but one cannot directly apply lasers
in that direction, so in practice both of the laser beams have a $45^{o}$ to
the trap axis \cite{Liebfried,Haljan}. However, for the TP\ quantum gate,
one can apply counter-propagating laser beams along the $x$-axis so that $%
\Delta \mathbf{k}$ is perpendicular to the ion string. Because of
this difference, the laser intensity ratio $I_{x}/I_{z}$ is actually
$\sqrt{\beta _{x}/2}$ instead of $\sqrt{\beta _{x}}$. The above
laser configuration difference also gives some practical advantages
for the TP\ quantum gates: first, as the laser beams are
perpendicular to the ion string, it is easy to have the same
relative laser phase for different ions, as assumed in the gates
above. For LP gates, it is difficult to achieve such a condition for
different ions as they are not equally spaced. It usually requires
subtle control of the ion distance through adjustment of the trap
frequency \cite{Sackett,Liebfried}, and such a technique is not
scalable to many ions. Second, with the transverse focused laser
beams, it is also easier to achieve separate addressing of different
ions. For the TP gate, one does not need to have a large
longitudinal $\omega _{z}$ to achieve the Lamb-Dicke condition, so
the ion distance is not limited, and one can combine separate
addressing with a many-ion setup, which is desired for scalable
quantum computation.

\textbf{Arbitrary-speed TP gates through minimal control of the laser beams}%
. As we have mentioned before, the TP\ modes have small frequency
splittings, so resolving a particular TP\ mode could be difficult
for a large ion array. If we want to achieve a high-speed gate, it
is necessary to go beyond the sideband addressing (single-mode)
limit. Fortunately, for any practical qubit number (up to a few
hundreds, for instance), it is always possible to take into account
of all the phonon modes and design
high-fidelity gates with no limitation to the gate speed \cite{f1,f2,Zhu2005}%
. These fast gates are in general based on control of the laser pulses.

Here, similar to Ref. \cite{Zhu2005}, we use a simple sequence of
laser pulses which take minimum experimental control. We chop a
continuous-wave laser beam into $m$ equal-time segments (through
acoustic/electric modulator, for instance), with a constant but controllable
Rabi frequency $\Omega _{p}$ for the $p$th ($p=1,2,\cdots ,m$) segment. The
state-dependent force $\chi(t)$ in the Hamiltonian (3) then takes the form $%
\chi(t)=\Omega _{p}\sin (\mu t)$ for the time interval $(p-1)\tau
/m\leq
t<p\tau /m$ (For simplicity of the notation, we omit the direction index $%
\xi $ in the following and take $\xi =x$ by default). With a large
number of ions but a small number of control parameters $\Omega
_{p}$, the displacements $\alpha _{j}^{k}(\tau )$ (and thus $\phi
_{j}(\tau )$) in the evolution operator (4) may not exactly reduce
to zero. But as long as they are small, we still can get a
high-fidelity gate. In this case, for a gate on the ion-pair $(j,n)$
in an $N$-ion array, the gate infidelity due to the residue small
displacements $\alpha _{j}^{k}$ is given by

\begin{equation}
F_{in}\approx \sum_{k}\bar{\beta}_{k}[|\alpha _{j}^{k}|^{2}+|\alpha
_{n}^{k}|^{2}]/4,
\end{equation}
where $\bar{\beta}_{k}=\coth [\sqrt{\mu _{k}/4}ln(1+1/\bar{n}_{1})]$ with $%
\bar{n}_{1}$ representing the mean phonon number of the CM TP mode. For the
above expression, we have assumed that the TP modes are in their thermal
equilibrium state. The task of control is to get a small infidelity $F_{in}$
by choosing an optimal laser detuning $\mu $ and a minimum number of the
control parameters $\Omega _{p}$.

\begin{figure}[tbph]
\label{Fig2} \includegraphics[height=3cm,width=8cm]{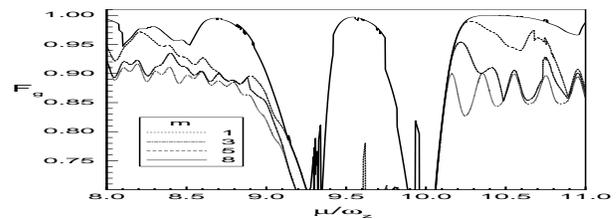}
\caption{For the two edge ions in a 10-ion array, the TP gate fidelity as a
function of the detuning $\protect\mu $ with the gate time $\protect\tau =5%
\protect\tau _{0}$,\ $\bar{n}_{1}=3$,\ $\protect\beta _{x}=10$, and the
number of segments $m=1,3,5,8$, respectively.}
\end{figure}

First, let us consider the gate with a single continuous-wave laser beam
(the number of segments $m=1$). The only control parameter is the detuning $%
\mu $. We find that for the gate on two edge ions (1st and 2nd ions) in a $10
$-ion array, as long as the gate time $\tau \geq 37\tau _{0}$, where $\tau
_{0}\equiv 2\pi /\omega _{z}$, the gate infidelity $F_{in}\leq 0.99\%$. For
this and the following calculations, we take $\bar{n}_{1}\approx 3$ which
corresponds to a pretty high temperature. The optimal $\mu $ is very close
to the value $\omega _{x}+2\pi /\tau $. Note that a gate with $\tau \approx
37\tau _{0}$ has been faster than any ion gate implemented so far in the lab
\cite{Liebfried,Haljan}. For this gate, the time $\tau $ is close to $\tau
_{\Delta }\equiv 2\pi /\Delta $ ($\tau \approx 1.8\tau _{\Delta }$), where $%
\Delta $ is the frequency splitting between the CM\ TP\ mode and the bending
mode. So, besides the dominant CM mode, various TP\ modes are indeed
slightly excited during the gate and contribute to the conditional phase $%
\phi _{jn}(\tau )$. But with the optimal $\mu $, all these modes evolve
along an almost-closed loop in the phase space ($\alpha _{i}^{k}\approx 0$,
although not exactly), so we still have a high fidelity gate. The required
Rabi frequency from the exact numerical calculation is very close to the one
given by the analytic formula (7) from the single-mode approximation (within
a $6\%$ error).

If we further increase the gate speed with $\tau <37\tau _{0}$, the gate
fidelity quickly decreases, so we need to chop the continuous-wave laser
beam into more segments with $m>1$ to increase the fidelity. With a
sufficiently large $m$, a high-fidelity gate with an arbitrary gate speed
can be achieved. In Fig.2, we show the calculation result for the gate time $%
\tau =5\tau _{0}$. With the number of segments $m=1,3,5,8$, the gate
infidelity is given by $10\%,\ 4.9\%,\ 1.0\%,\ 0.1\%$, respectively, with
the optimized parameters $\mu $ and $\Omega _{1},\Omega _{2},\cdots ,\
\Omega _{m}$. We also calculate the gate infidelity for other ion-pairs, and
the results are qualitatively similar. For instance, with $\tau =5\tau _{0}$
and the number of segments $m=1,3,5,8$, respectively, the gate infidelity $%
F_{in}$ is given by $5.5\%,\ 1.8\%,\ 0.22\%,$ $0.07\%$ for the two center
ions (5th and 6th ions), and by $40\%,\ 25\%,\ 8.5\%,\ 0.99\%$ for the 1st
and 10th ions at the far ends of the string (the worst case).

In summary, we have proposed to use the transverse phonon mode to achieve
quantum gates, and shown that these TP gates are much less sensitive to
ion heating and thermal motion compared with conventional gates
based on the longitudinal modes. We also show how to achieve arbitrary-speed
TP quantum gates based on control of the laser pulses.

This work is supported by the NSA and the DTO under ARO contract
W911NF-04-1-0234, the NSF under grant 0431476, and the A. P. Sloan
Foundation. S.L.Z. also acknowledges support by the NCET.


\begin{thebibliography}{99}
\bibitem{Fidelity}  The gate fidelity is defined as $F_{g}^{\xi }=\langle
\Psi _{0}|U_{ij}^{\dagger }\rho _{r}U_{ij}|\Psi _{0}\rangle $ with a
typical initial state $|\Psi _{0}\rangle =(|0\rangle _{i}+|1\rangle
_{j})\otimes (|0\rangle _{j}+|1\rangle _{j})/2$. The density
operator $\rho _{r}=Tr_{m}[U_{\xi }(\tau )|\Psi _{0}\rangle \langle
\Psi _{0}|U_{\xi }^{\dagger }(\tau )]$, where the trace is over the
motional states of all the ions.
\bibitem{note}  For this calculation, the gate time $\tau $ remains the same
as it is fixed by the condition $\phi _{j}^{\xi }(\tau )=\phi
_{n}^{\xi }(\tau )=0$.
\end{thebibliography}
\end{document}